\catcode`\@=11
\expandafter\ifx\csname @iasmacros\endcsname\relax
	\global\let\@iasmacros=\par
\else	\immediate\write16{}
	\immediate\write16{Warning:}
	\immediate\write16{You have tried to input iasmacros more than once.}
	\immediate\write16{}
	\endinput
\fi
\catcode`\@=12


\def\rmb{\seventeenrm}


\def\halfspace{\baselineskip=1.5\normalbaselineskip}
\def\doublespace{\baselineskip=2\normalbaselineskip}


\def\AB{\bigskip\parindent=40pt
        \centerline{\bf ABSTRACT}\medskip\halfspace\narrower}
\def\AE{\bigskip\nonarrower\doublespace}
\def\nonarrower{\advance\leftskip by-\parindent
	\advance\rightskip by-\parindent}


\def\boxit#1{\vbox{\hrule\hbox{\vrule\kern3pt
	\vbox{\kern3pt#1\kern3pt}\kern3pt\vrule}\hrule}}

\def\hence{\leavevmode\hbox{\bf .\raise5.5pt\hbox{.}.} }

\def\dalemb#1#2{{\vbox{\hrule height.#2pt
	\hbox{\vrule width.#2pt height#1pt \kern#1pt \vrule width.#2pt}
	\hrule height.#2pt}}}
\def\gtorder{\mathrel{\raise.3ex\hbox{$>$}\mkern-14mu
             \lower0.6ex\hbox{$\sim$}}}
\def\ltorder{\mathrel{\raise.3ex\hbox{$<$}\mkern-14mu
             \lower0.6ex\hbox{$\sim$}}}

\newdimen\fullhsize
\newbox\leftcolumn
\def\twoup{\hoffset=-.5in \voffset=-.25in
  \hsize=4.75in \fullhsize=10in \vsize=6.9in
  \def\fullline{\hbox to\fullhsize}
  \let\lr=L
  \output={\if L\lr
        \global\setbox\leftcolumn=\columnbox\global\let\lr=R \advancepageno
      \else \doubleformat \global\let\lr=L\fi
    \ifnum\outputpenalty>-20000 \else\dosupereject\fi}
  \def\doubleformat{\shipout\vbox{
    \fullline{\box\leftcolumn\hfil\columnbox}\advancepageno}}
  \def\columnbox{\leftline{\vbox{\makeheadline\pagebody\makefootline}}}
  \tolerance=1000 }
\catcode`\@=11					



\font\fiverm=cmr5				
\font\fivemi=cmmi5				
\font\fivesy=cmsy5				
\font\fivebf=cmbx5				

\skewchar\fivemi='177
\skewchar\fivesy='60


\font\sixrm=cmr6				
\font\sixi=cmmi6				
\font\sixsy=cmsy6				
\font\sixbf=cmbx6				

\skewchar\sixi='177
\skewchar\sixsy='60


\font\sevenrm=cmr7				
\font\seveni=cmmi7				
\font\sevensy=cmsy7				
\font\sevenit=cmti7				
\font\sevenbf=cmbx7				

\skewchar\seveni='177
\skewchar\sevensy='60


\font\eightrm=cmr8				
\font\eighti=cmmi8				
\font\eightsy=cmsy8				
\font\eightit=cmti8				
\font\eightbf=cmbx8				

\skewchar\eighti='177
\skewchar\eightsy='60


\font\ninei=cmmi9
\font\ninesy=cmsy9

\skewchar\ninei='177
\skewchar\ninesy='60


\font\tenrm=cmr10				
\font\teni=cmmi10				
\font\tensy=cmsy10				
\font\tenex=cmex10				
\font\tenit=cmti10				
\font\tensl=cmsl10				
\font\tenbf=cmbx10				
\font\tentt=cmtt10				
\font\tenss=cmss10				
\font\tensc=cmcsc10				
\font\tenbi=cmmib10				

\skewchar\teni='177
\skewchar\tenbi='177
\skewchar\tensy='60

\def\tenpoint{\ifmmode\err@badsizechange\else
	\textfont0=\tenrm \scriptfont0=\sevenrm \scriptscriptfont0=\fiverm
	\textfont1=\teni  \scriptfont1=\seveni  \scriptscriptfont1=\fivemi
	\textfont2=\tensy \scriptfont2=\sevensy \scriptscriptfont2=\fivesy
	\textfont3=\tenex \scriptfont3=\tenex   \scriptscriptfont3=\tenex
	\textfont4=\tenit \scriptfont4=\sevenit \scriptscriptfont4=\sevenit
	\textfont5=\tensl
	\textfont6=\tenbf \scriptfont6=\sevenbf \scriptscriptfont6=\fivebf
	\textfont7=\tentt
	\textfont8=\tenbi \scriptfont8=\seveni  \scriptscriptfont8=\fivemi
	\def\rm{\tenrm\fam=0 }%
	\def\it{\tenit\fam=4 }%
	\def\sl{\tensl\fam=5 }%
	\def\bf{\tenbf\fam=6 }%
	\def\tt{\tentt\fam=7 }%
	\def\ss{\tenss}%
	\def\sc{\tensc}%
	\def\bmit{\fam=8 }%
	\rm\setparameters\setbaselines\fi}


\font\twelverm=cmr12				
\font\twelvei=cmmi12				
\font\twelvesy=cmsy10	scaled\magstep1		
\font\twelveex=cmex10	scaled\magstep1		
\font\twelveit=cmti12				
\font\twelvesl=cmsl12				
\font\twelvebf=cmbx12				
\font\twelvett=cmtt12				
\font\twelvess=cmss12				
\font\twelvesc=cmcsc10	scaled\magstep1		
\font\twelvebi=cmmib10	scaled\magstep1		

\skewchar\twelvei='177
\skewchar\twelvebi='177
\skewchar\twelvesy='60

\def\twelvepoint{\ifmmode\err@badsizechange\else
	\textfont0=\twelverm \scriptfont0=\eightrm \scriptscriptfont0=\sixrm
	\textfont1=\twelvei  \scriptfont1=\eighti  \scriptscriptfont1=\sixi
	\textfont2=\twelvesy \scriptfont2=\eightsy \scriptscriptfont2=\sixsy
	\textfont3=\twelveex \scriptfont3=\tenex   \scriptscriptfont3=\tenex
	\textfont4=\twelveit \scriptfont4=\eightit \scriptscriptfont4=\sevenit
	\textfont5=\twelvesl
	\textfont6=\twelvebf \scriptfont6=\eightbf \scriptscriptfont6=\sixbf
	\textfont7=\twelvett
	\textfont8=\twelvebi \scriptfont8=\eighti  \scriptscriptfont8=\sixi
	\def\rm{\twelverm\fam=0 }%
	\def\it{\twelveit\fam=4 }%
	\def\sl{\twelvesl\fam=5 }%
	\def\bf{\twelvebf\fam=6 }%
	\def\tt{\twelvett\fam=7 }%
	\def\ss{\twelvess}%
	\def\sc{\twelvesc}%
	\def\bmit{\fam=8 }%
	\rm\setparameters\setbaselines\fi}


\font\fourteenrm=cmr12	scaled\magstep1		
\font\fourteeni=cmmi12	scaled\magstep1		
\font\fourteensy=cmsy10	scaled\magstep2		
\font\fourteenex=cmex10	scaled\magstep2		
\font\fourteenit=cmti12	scaled\magstep1		
\font\fourteensl=cmsl12	scaled\magstep1		
\font\fourteenbf=cmbx12	scaled\magstep1		
\font\fourteentt=cmtt12	scaled\magstep1		
\font\fourteenss=cmss12	scaled\magstep1		
\font\fourteensc=cmcsc10 scaled\magstep2	
\font\fourteenbi=cmmib10 scaled\magstep2	

\skewchar\fourteeni='177
\skewchar\fourteenbi='177
\skewchar\fourteensy='60

\def\fourteenpoint{\ifmmode\err@badsizechange\else
	\textfont0=\fourteenrm \scriptfont0=\tenrm \scriptscriptfont0=\sevenrm
	\textfont1=\fourteeni  \scriptfont1=\teni  \scriptscriptfont1=\seveni
	\textfont2=\fourteensy \scriptfont2=\tensy \scriptscriptfont2=\sevensy
	\textfont3=\fourteenex \scriptfont3=\tenex \scriptscriptfont3=\tenex
	\textfont4=\fourteenit \scriptfont4=\tenit \scriptscriptfont4=\sevenit
	\textfont5=\fourteensl
	\textfont6=\fourteenbf \scriptfont6=\tenbf \scriptscriptfont6=\sevenbf
	\textfont7=\fourteentt
	\textfont8=\fourteenbi \scriptfont8=\tenbi \scriptscriptfont8=\seveni
	\def\rm{\fourteenrm\fam=0 }%
	\def\it{\fourteenit\fam=4 }%
	\def\sl{\fourteensl\fam=5 }%
	\def\bf{\fourteenbf\fam=6 }%
	\def\tt{\fourteentt\fam=7}%
	\def\ss{\fourteenss}%
	\def\sc{\fourteensc}%
	\def\bmit{\fam=8 }%
	\rm\setparameters\setbaselines\fi}


\font\seventeenrm=cmr10 scaled\magstep3		


\newdimen\rp@
\newcount\@basestretchnum
\newskip\@baseskip
\newskip\headskip
\newskip\footskip


\def\setparameters{\rp@=.1em
	\headskip=24\rp@
	\footskip=\headskip
	\delimitershortfall=5\rp@
	\nulldelimiterspace=1.2\rp@
	\scriptspace=0.5\rp@
	\abovedisplayskip=10\rp@ plus3\rp@ minus5\rp@
	\belowdisplayskip=10\rp@ plus3\rp@ minus5\rp@
	\abovedisplayshortskip=5\rp@ plus2\rp@ minus4\rp@
	\belowdisplayshortskip=10\rp@ plus3\rp@ minus5\rp@
	\normallineskip=\rp@
	\lineskip=\normallineskip
	\normallineskiplimit=0pt
	\lineskiplimit=\normallineskiplimit
	\jot=3\rp@
	\setbox0=\hbox{\the\textfont3 B}\p@renwd=\wd0
	\skip\footins=12\rp@ plus3\rp@ minus3\rp@
	\skip\topins=0pt plus0pt minus0pt}


\def\setbaselines{\maxdepth=4\rp@\baselinestretch=\@basestretchnum}


\def\baselinestretch{\afterassignment\@basestretch\@basestretchnum}
\def\@basestretch{%
	\@baseskip=12\rp@ \divide\@baseskip by1000
	\normalbaselineskip=\@basestretchnum\@baseskip
	\baselineskip=\normalbaselineskip
	\bigskipamount=\the\baselineskip
		plus.25\baselineskip minus.25\baselineskip
	\medskipamount=.5\baselineskip
		plus.125\baselineskip minus.125\baselineskip
	\smallskipamount=.25\baselineskip
		plus.0625\baselineskip minus.0625\baselineskip
	\setbox\strutbox=\hbox{\vrule height.708\baselineskip
		depth.292\baselineskip width0pt }}



\def\makeheadline{\vbox to0pt{\baselinestretch=1000
	\vskip-\headskip \vskip1.5pt
	\line{\vbox to\ht\strutbox{}\the\headline}\vss}\nointerlineskip}

\def\makefootline{\baselineskip=\footskip\line{\the\footline}}

\def\big#1{{\hbox{$\left#1\vbox to8.5\rp@ {}\right.\n@space$}}}
\def\Big#1{{\hbox{$\left#1\vbox to11.5\rp@ {}\right.\n@space$}}}
\def\bigg#1{{\hbox{$\left#1\vbox to14.5\rp@ {}\right.\n@space$}}}
\def\Bigg#1{{\hbox{$\left#1\vbox to17.5\rp@ {}\right.\n@space$}}}


\mathchardef\alpha="710B
\mathchardef\beta="710C
\mathchardef\gamma="710D
\mathchardef\delta="710E
\mathchardef\epsilon="710F
\mathchardef\zeta="7110
\mathchardef\eta="7111
\mathchardef\theta="7112
\mathchardef\iota="7113
\mathchardef\kappa="7114
\mathchardef\lambda="7115
\mathchardef\mu="7116
\mathchardef\nu="7117
\mathchardef\xi="7118
\mathchardef\pi="7119
\mathchardef\rho="711A
\mathchardef\sigma="711B
\mathchardef\tau="711C
\mathchardef\upsilon="711D
\mathchardef\phi="711E
\mathchardef\chi="711F
\mathchardef\psi="7120
\mathchardef\omega="7121
\mathchardef\varepsilon="7122
\mathchardef\vartheta="7123
\mathchardef\varpi="7124
\mathchardef\varrho="7125
\mathchardef\varsigma="7126
\mathchardef\varphi="7127
\mathchardef\imath="717B
\mathchardef\jmath="717C
\mathchardef\ell="7160
\mathchardef\wp="717D
\mathchardef\partial="7140
\mathchardef\flat="715B
\mathchardef\natural="715C
\mathchardef\sharp="715D


\def\err@badsizechange{%
	\immediate\write16{--> Size change not allowed in math mode, ignored}}

\baselinestretch=1000
\tenpoint

\catcode`\@=12					

\twelvepoint
\doublespace
\font\itb=cmr10 scaled 1440
{\nopagenumbers{
\rightline{IASSNS-HEP-96/89}
\rightline{~~~September, 1996}
\bigskip\bigskip
\centerline{\rmb Nonadiabatic Geometric Phase}
\centerline{\rmb in Quaternionic Hilbert Space}
\medskip
\centerline{\itb Stephen L. Adler}
\centerline{\bf Institute for Advanced Study}
\centerline{\bf Princeton, NJ 08540}
\centerline{\it adler@sns.ias.edu}
\medskip
\medskip
\centerline{\itb Jeeva Anandan}
\centerline{\bf Department of Physics and Astronomy, University of  
South Carolina}
\centerline{\bf Columbia, South Carolina 29208}
\centerline{\it jeeva@nuc003.psc.sc.edu}
\medskip

\bigskip\bigskip
\leftline{To appear in a special issue of FOUNDATIONS OF PHYSICS dedicated}
\leftline{to Professor Lawrence Horwitz on the occasion of his 65th birthday.}
\bigskip\bigskip
}}
\AB
We develop the theory of the nonadiabatic geometric phase, in both the 
Abelian and non-Abelian cases, in quaternionic Hilbert space.
\AE
\bigskip\bigskip
\vfill\eject
\pageno=2
\centerline{{\bf 1.~~Introduction}}
\bigskip
The theory of geometric phases associated with cyclic evolutions of a 
physical system is now a well-developed subject in complex Hilbert space.
The seminal work of Berry on the adiabatic single state (Abelian) case [1] 
has been extended to the non-Abelian case of the adiabatic evolution of
a set of degenerate states [2], and both of these 
have been further extended [3, 4] 
to show that there is a geometric phase associated with any cyclic but 
nonadiabatic evolution of a single quantum state or of a degenerate group 
of quantum states.  

In this paper we take up another direction for generalization of the 
geometric phase, from quantum mechanics in complex Hilbert 
space to quantum mechanics [5, 6] in quaternionic Hilbert space.  
The generalization of the adiabatic geometric phase to quaternionic Hilbert 
space was given in Ref. 6, where it was shown that for states of nonzero 
energy the adiabatic geometric phase is complex, as opposed to quaternionic, 
with a 
quaternionic adiabatic geometric phase occurring only for the adiabatic 
cyclic evolution of zero energy states.  Consideration of nonadiabatic 
cyclic evolutions was also begun in Ref. 6, but the discussion given there 
is incomplete.  While Sec. 5.8 of Ref.~6 constructed a nonadiabatic cyclic 
invariant phase, it did not address the 
problem of separating this phase into a dynamical part determined by the 
quantum mechanical Hamiltonian, and a geometric part that depends only 
on the ray orbit and is independent of the Hamiltonian.  

The purpose of the present paper is to give a complete discussion of the 
nonadiabatic geometric phase in quaternionic Hilbert space.  In Sec.~2 
we give a very brief survey of the properties of quantum mechanics 
in quaternionic Hilbert space that are needed in the analysis that follows.    
In Sec.~3 we consider the cyclic nonadiabatic evolution of a single quantum 
state, and show how to explictly generalize to quaternionic Hilbert space 
the construction of a nonadiabatic geometric phase given in Ref. 3.  
In Sec.~4 we extend our analysis to the case of a degenerate group of  
states, thereby obtaining a quaternionic nonadiabatic non-Abelian geometric 
phase corresponding to the complex construction given in Ref. 4.
A brief summary and  discussion of our results is given in Sec.~5.  
\bigskip
\centerline{{\bf 2.  Quantum Mechanics in Quaternionic Hilbert Space}}
\bigskip
Only a few properties of quaternionic quantum mechanics are needed for 
the discussion that follows; the reader wishing to learn more than we 
can present here should consult Ref.~6.  In quaternionic quantum mechanics, 
the Dirac transition amplitudes $\langle \psi | \phi \rangle$ are 
quaternion valued, that is, they have the form
$$\langle \psi | \phi \rangle=r_0+r_1 i+r_2 j+r_3 k~,\eqno(1)$$
where $r_{0,1,2,3}$ are real numbers and where $i,j,k$ are quaternion 
imaginary units obeying the associative algebra 
$i^2=j^2=k^2=-1$ and $ij=-ji=k,~jk=-kj=i,~ki=-ik=j$.  Because quaternion 
multiplication is noncommutative, two independent Dirac transition 
amplitudes $\langle \psi | \phi \rangle $ and $\langle \kappa | \eta \rangle$ 
in general do not commute with one another, unlike the situatation in 
standard complex quantum mechanics, where all Dirac transition amplitudes are 
complex numbers and mutually commute.  The transition probability  
corresponding to the amplitude of Eq.~(1) is given by 
$$P(\psi,\phi)=|\langle \psi | \phi \rangle|^2\equiv 
\overline{\langle \psi | \phi \rangle}\langle \psi | \phi \rangle
=r_0^2+r_1^2+r_2^2+r_3^2~,\eqno(2)$$
where the bar denotes the quaternion conjugation operation $\{i,j,k\} \to 
\{-i,-j,-k\}$ and where we have assumed the states $|\psi \rangle$ and 
$|\phi\rangle$ to be unit normalized.  
Since the quaternion norm defined by Eq.~(2) has the multiplicative 
norm property 
$$|q_1 q_2|=|q_1|~|q_2|~,\eqno(3)$$
the transition probability of Eq.~(2) is unchanged when the state vector 
$| \phi \rangle$  is right multiplied by a quaternion $\omega$ of unit  
magnitude, 
$$|\phi\rangle \to |\phi \rangle \omega,~~|\omega|=1~~
\Rightarrow P(\psi,\phi) \to P(\psi, \phi)
~.\eqno(4)$$
Hence as in complex quantum mechanics, physical states are associated with 
Hilbert space rays of the form $\{|\phi\rangle \omega~: |\omega|=1\}$, and 
the transition probability of Eq.~(2) is the same for any ray representative 
state vectors $|\psi \rangle $ and $|\phi \rangle$ chosen from their 
corresponding rays.  In the next section, we shall follow Ref.~3 in denoting 
quaternionic Hilbert space by ${\cal H}$,  and the projective Hilbert 
space of rays of ${\cal H}$ by ${\cal P}$.

Time evolution of the state vector $| \psi \rangle$  is described in 
quaternionic quantum mechanics by the Schr\"odinger equation 
$${\partial |\psi\rangle \over \partial t}=-\tilde H |\psi \rangle~,
\eqno(5a)$$
with 
$$\tilde H=-\tilde H^{\dagger}~\eqno(5b)$$
an anti-self-adjoint Hamiltonian.  From Eqs.~(5a,~b) we see that the 
Dirac transition amplitude $\langle \psi |\phi \rangle$ is time independent, 
$$\eqalign{
{\partial \over \partial t} \langle \psi| \phi \rangle=& 
({\partial \over \partial t} \langle \psi|)| \phi \rangle+ 
\langle \psi| {\partial \over \partial t} | \phi \rangle \cr 
=&\langle \psi| \tilde H - \tilde H |\phi \rangle=0~,\cr
}\eqno(6)$$
and thus the Schr\"odinger dynamics of state vectors 
preserves the inner product structure of Hilbert space.  
The dynamics of Eqs.~(5,~6) is evidently preserved under right linear 
superposition of states with quaternionic constants, 
$$\eqalign{
{\partial |\psi\rangle \over \partial t}=&-\tilde H |\psi \rangle~,
{\partial |\phi\rangle \over \partial t}=-\tilde H |\phi \rangle ~~\Rightarrow 
\cr
{\partial(|\psi\rangle q_1+|\phi\rangle q_2) \over \partial t}=&
-\tilde H (|\psi\rangle q_1+|\phi\rangle q_2)~.\cr
}\eqno(7)$$
Equation (7) illustrates two general features of our conventions for 
quaternionic quantum mechanics,  
which are that linear operators (such as $\tilde H$) act on Hilbert space 
state vectors by 
multiplication from the left, whereas quaternionic numbers (the scalars 
of Hilbert space) act on state vectors by multiplication from the right.   
Adherence to these ordering conventions is essential because of the 
noncommutative nature of quaternionic multiplication.  
\bigskip
\centerline{\bf 3.  The Nonadiabatic Abelian Quaternionic Geometric Phase}
\bigskip
Let us now consider a unit normalized quaternionic Hilbert space 
state $| \psi(t) \rangle$
which undergoes a cyclic evolution between the times $t=0$ and $t=T$.  Since 
physical states are associated with rays, this means that 
$$|\psi(T)\rangle=  |\psi(0) \rangle \Omega~,~~|\Omega|=1~, \eqno(8)$$
and so the orbit ${\cal C}$ of $|\psi(t) \rangle$ in ${\cal H}$ projects to 
a closed curve  $\hat{\cal C}$ in the projective Hilbert space ${\cal P}$.  

Let us now define a state $|\hat {\psi}(t) \rangle$ that is equal to 
$|\psi(t) \rangle$ at $t=0$, that differs from $|\psi(t) \rangle$ only by   
a reraying at general times, i.e.,  
$$\eqalign{
|\psi(t)\rangle=&|\hat{\psi}(t)\rangle \hat{\omega}(t)~,\cr
|\hat{\omega}(t)|=&1~, \cr
\hat{\omega}(0)=1~,\cr
}\eqno(9a)$$
and that evolves in time by parallel transport, 
i.e., 
$$\langle \hat{\psi}(t) | {\partial | \hat{\psi}(t) \rangle 
\over \partial t} 
=0~.\eqno(9b)$$
The conditions of Eqs.~(9a,~b) uniquely determine $\hat{\omega}(t)$, and 
hence the state $|\hat{\psi}(t)\rangle$, as follows.  Substituting 
the first line of Eq.~(9a) into the Schr\"odinger equation of Eq.~(5a), 
we get
$$\eqalign{
-\tilde H |\hat{\psi}(t)\rangle \hat{\omega}(t)=&
-\tilde H |\psi(t)\rangle\cr
=&{\partial |\psi(t)\rangle \over \partial t} 
=|\hat{\psi}(t)\rangle {d \hat{\omega}(t) \over d t}
+{\partial |\hat{\psi}(t)\rangle \over \partial t} \hat{\omega}(t)~.\cr
}\eqno(10)$$
Taking the inner product of this equation with the state $\langle \hat{\psi}
(t)|$, and using the 
unit normalization of the state vector $|\hat{\psi}(t)\rangle$ together
with the parallel transport condition of Eq.~(9b), we get
$${d \hat{\omega}(t) \over d t}=
-\langle \hat{\psi}(t)|\tilde H|\hat{\psi}(t)\rangle 
\hat{\omega}(t)~.\eqno(11)$$
This differential equation can be immediately integrated to give 
$$\hat{\omega}(t)=T_{\ell}e^{-\int_0^t dv \langle \hat{\psi}(v)|\tilde H
|\hat{\psi}(v) \rangle}~,\eqno(12)$$ 
where $T_{\ell}$ denotes the time ordered product which orders later times to 
the 
left, and where we have used the initial condition on the third line of 
Eq.~(9a).  In particular, Eq.~(12) gives us a formula for the value 
$\hat{\omega}(T)$ at the end of the cyclic evolution.  We shall see that 
this has the interpretation of the dynamics-dependent part of the total 
phase change $\Omega$.

To relate Eq.~(12) to the total phase change, we use Eqs.~(8) and (9a) to 
write 
$$|\hat{\psi}(T)\rangle \hat{\omega}(T)= |\psi(T) \rangle=
|\psi(0) \rangle \Omega=|\hat{\psi}(0)\rangle \Omega~,\eqno(13a)$$
so that taking the inner product with $\langle \hat{\psi}(0)|$ gives
$$\Omega=\langle \hat{\psi}(0)|\hat{\psi}(T) \rangle 
\hat{\omega}(T)~.\eqno(13b)$$
To complete the calculation, we must now evaluate the inner product 
appearing in Eq.~(13b).  To do this, we introduce a third state vector 
$|\tilde{\psi}(t)\rangle$ which differs from $|\hat{\psi}(t)\rangle$ by a 
change of ray representative, by writing
$$\eqalign{
|\hat{\psi}(t)\rangle =& |\tilde{\psi}(t)\rangle \tilde{\omega}(t)~,\cr
|\tilde{\omega}(t)|=&1~,\cr
\tilde{\omega}(0)=&1~,\cr
}\eqno(14a)$$
and by requiring that $\tilde{\psi}$ should be continuous over the orbit 
${\cal C}$, 
$$|\tilde{\psi}(T)\rangle=|\tilde{\psi}(0)\rangle~.\eqno(14b)$$
Differentiating the first line of Eq.~(14a) with respect to time, we get 
$${\partial |\hat{\psi}(t)\rangle \over \partial t}=
{\partial |\tilde{\psi}(t)\rangle \over \partial t} \tilde{\omega}(t)
+|\tilde{\psi}(t)\rangle {d \tilde{\omega}(t) \over d t}~.
\eqno(15)$$
Taking the inner product of Eq.~(15) with 
$\tilde {\omega}(t)\langle \hat{\psi}(t)|$, 
using the parallel transport condition of Eq.~(9b) together with the 
first line of Eq.~(14a), and abbreviating the time derivative ${\partial \over 
\partial t}$ by a dot, we obtain
$$0=\langle \tilde{\psi}(t)| \dot{\tilde{\psi}}(t) \rangle \tilde{\omega}(t)
+\langle \tilde{\psi}(t)|\tilde{\psi}(t)\rangle \dot{\tilde{\omega}}(t)~.
\eqno(16a)$$
Since the second line of Eq.~(14a) implies that the state $|\tilde{\psi}(t)
\rangle$ is unit normalized, Eq.~(16a) simplifies to 
$$\dot{\tilde{\omega}}(t)=
-\langle \tilde{\psi}(t)| \dot{\tilde{\psi}}(t) \rangle \tilde{\omega}(t)~,
\eqno(16b)$$
which can be immediately integrated to give 
$$\tilde{\omega}(t)=T_{\ell}e^{-\int_0^t dv \langle \tilde{\psi}(v)|
\dot{\tilde{\psi}}(v) \rangle}~,\eqno(17)$$ 
with $T_{\ell}$ as before indicating a time ordered product.  In particular, 
Eq.~(17) gives us a formula for $\tilde{\omega}(T)$.  But from Eqs.~(14a, b) 
we have
$$|\hat{\psi}(T)\rangle=|\tilde{\psi}(T)\rangle \tilde{\omega}(T)=
|\tilde{\psi}(0)\rangle \tilde{\omega}(T)=|\hat{\psi}(0)\rangle 
\tilde{\omega}(T)~,\eqno(18a)$$
and so taking the inner product of Eq.~(18a) with $\langle \hat{\psi}(0)|$
we get 
$$\langle \hat{\psi}(0)|\hat{\psi}(T)\rangle = \tilde{\omega}(T)~,\eqno(18b)$$
determining the inner product appearing in Eq.~(13b).  

We thus get as our final result, 
$$\Omega=\Omega_{\rm geometric} \Omega_{\rm dynamical}~,\eqno(19a)$$
with 
$$\Omega_{\rm geometric}\equiv\tilde{\omega}(T)=
T_{\ell}e^{-\int_0^T dv \langle \tilde{\psi}(v)|
\dot{\tilde{\psi}}(v) \rangle}~,\eqno(19b)$$ 
and with
$$\Omega_{\rm dynamical}\equiv\hat{\omega}(T)=
T_{\ell}e^{-\int_0^T dv \langle \hat{\psi}(v)|\tilde H
|\hat{\psi}(v) \rangle}~.\eqno(19c)$$ 
The dynamical part of the phase is so called because it depends explicitly  
on $\tilde H$, as well as on the orbit $\hat{\cal C}$ in the projective 
Hilbert space $\cal P$; it is uniquely determined   
by the conditions of Eqs.~(9a, b), since these conditions uniquely   
determine the state $|\hat{\psi}(t)\rangle$.
The geometric part of the phase is so called because, as we shall now 
show, it depends uniquely on the projective orbit $\hat{\cal C}$ 
up to an overall quaternion automorphism transformation.  To see this, 
let us make the reraying 
$$|\tilde{\psi}(t)\rangle \to  |\tilde{\psi}^{\,\prime}\rangle 
\omega^{\,\prime}
(t),~~|\omega^{\,\prime}|=1~, \eqno(20a)$$
with $\omega^{\,\prime}(t)$ continuous over the orbit $\cal C$ so that 
$$\omega^{\,\prime}(T)=\omega^{\,\prime}(0)~.\eqno(20b)$$
Then (as shown in detail in Sec. 5.8 of Ref.~6) 
the properties of the time ordered integral in Eq.~(19b) imply that under 
this transformation, 
$$\Omega_{\rm geometric} \to \overline{\omega}^{\,\prime}(T) \Omega_{\rm 
geometric}
\,\omega^{\,\prime}(0)~,\eqno(21a)$$
which by the continuity condition of Eq.~(20b) reduces to the 
quaternion automorphism transformation 
$$\Omega_{\rm geometric} \to \overline{\omega}^{\,\prime}(0) \Omega_{\rm 
geometric}
\,\omega^{\,\prime}(0)~.\eqno(21b)$$
Since for any two quaternions $q_1, q_2$ we have ${\rm Re}\, q_1 q_2=
{\rm Re}\, q_2 q_1$, with Re denoting the real part, Eq.~(21b) implies 
that 
$$\cos \gamma_{\rm geometric} \equiv {\rm Re} \Omega_{\rm 
geometric}~\eqno(22)$$ 
is a reraying invariant, and thus $\gamma_{\rm geometric}$ is a nonadiabatic 
geometric  phase angle that is a property solely of the projective 
orbit $\hat{\cal C}$.  The fact that the nonadiabatic geometric phase 
in quaternionic Hilbert space is only determined modulo $\pi$ is a reflection 
of the fact that $e^{i\gamma}$ is changed to $e^{-i\gamma}$ by 
the quaternion automorphism transformation 
$$e^{-i\gamma}=\bar j e^{i \gamma} j ~.\eqno(23)$$
Thus, to recover the result that the complex nonadiabatic geometric phase is 
determined modulo $2 \pi$ by embedding a complex Hilbert space in a 
quaternionic one and using Eqs.~(19a-c), one must exclude the possibility 
of making intrinsically quaternionic automorphism transformations involving 
the quaternion units $j$ or $k$, as in Eq.~(23).

In geometric terms, $\Omega_{\rm geometric}$ is the holonomy transformation 
of the connection $A \equiv \langle \tilde \psi | d \tilde \psi \rangle$.  
But since this connection is quaternion-imaginary valued, it is analogous 
to an $SO(3)$ gauge potential.  Therefore, the corresponding curvature 
is of the Yang-Mills type and is given by $F=dA + A \wedge A$.
  
An alternative expression for the total phase change $\Omega$ can be 
obtained [7] by writing
$$\eqalign{
|\psi(t)\rangle=&|\tilde{\psi}(t)\rangle \tilde {\chi}(t)~,\cr
\tilde{\chi}(t)=&\hat{\omega}(t)\tilde{\omega}(t)~,~~\tilde{\chi}(0)=1~.\cr
}\eqno(24)$$
Substituting Eq.~(24) into the Schr\"odinger equation and then 
taking the inner product with $\langle \tilde{\psi}(t)|$, we obtain
$${d \tilde{\chi}(t) \over dt}=-(\langle \tilde{\psi}(t) | \tilde H | 
\tilde{\psi}(t) \rangle + \langle \tilde{\psi}(t)|\dot{\tilde{\psi}}(t)
\rangle )\tilde{\chi}(t)~, \eqno(25a)$$
which can be integrated from $0$ to $T$ to give 
$$\Omega=T_{\ell}e^{-\int_0^T dv
(\langle \tilde{\psi}(v) | \tilde H | 
\tilde{\psi}(v) \rangle + \langle \tilde{\psi}(v)|\dot{\tilde{\psi}}(v)
\rangle)}~.\eqno(25b)$$
This procedure and the resulting formula of Eq.~(25b) are direct analogs 
of the derivation given in Ref.~3 for the complex Hilbert space case, but in 
quaternionic 
Hilbert space the two terms in the exponential are noncommutative, and so 
the exponential in Eq.~(25b) cannot be immediately factored into
dynamical and geometric phase factors.  As we have seen, to achieve this 
factorization it is necessary to use a two-step procedure, involving 
the parallel transported state $|\hat{\psi}(t)\rangle$ as well as the 
state $|\tilde{\psi}(t)\rangle$ that is continuous over the cycle.
\bigskip
\centerline{\bf 4.  The Nonadiabatic Non-Abelian Quaternionic Geometric Phase}
\bigskip                                                                 
We turn next to the quaternionic Hilbert space generalization of the 
complex nonadiabatic [4] non-Abelian [2] geometric phase.  We consider 
now a cyclic evolution in a $n$-dimensional Hilbert subspace $V_n$, i.e., 
$V_n(T)=V_n(0)$.  Let $|\psi_a(t)\rangle,~a=1,...,n$ be a complete orthonormal 
basis for $V_n$, so that the reraying invariant 
projection operator for $V_n$ is 
$$\rho_n(t)=\sum_{a=1}^n |\psi_a(t)\rangle \langle \psi_a(t)|~,\eqno(26a)$$
in terms of which the cyclic evolution condition takes the 
form 
$$\rho_n(T)=\rho_n(0)~.\eqno(26b)$$
Expressed in terms of the state vectors of the basis, the invariance of $V_n$
implies that the basis element $|\psi_b(T)\rangle$ must be a superposition  
of the basis elements $|\psi_a(0)\rangle$, multiplied from the right 
by quaternionic coefficients $U_{ab}$, 
$$|\psi_b(T)\rangle = \sum_{a=1}^n |\psi_a(0)\rangle U_{ab}~. \eqno(27)$$
Substituting Eq.~(27) into Eqs.~(26a, b), we find that invariance of the 
projection operator requires  
$$\eqalign{
\rho_n(T)=&\sum_{b=1}^n|\psi_b(T)\rangle \langle \psi_b(T)|  \cr
=&\sum_{a,b,c=1}^n |\psi_a(0)\rangle U_{ab} \overline{U}_{cb} \langle 
\psi_c(0)| \cr
=&\sum_{a=1}^n |\psi_a(0)\rangle \langle \psi_a(0)| = \rho_n(0)~, \cr
}\eqno(28a)$$
which implies that 
$$\sum_{b=1}^n U_{ab} \overline{U}_{cb}=\delta_{ac}~.\eqno(28b)$$
Similarly, orthonormality of the basis at $t=0$ and $t=T$ implies that 
$$\eqalign{
\delta_{ab}=&\langle \psi_a(T)|\psi_b(T) \rangle =\sum_{c,d=1}^n 
\overline{U}_{ca}\langle \psi_c(0)|\psi_d(0) \rangle U_{db} \cr  
=&\sum_{c,d=1}^n \overline{U}_{ca} \delta_{cd} U_{db}   
=\sum_{d=1}^n    \overline{U}_{da}  U_{db} ~.\cr
}\eqno(28c)$$
Hence $U$ is a $n \times n$ quaternion unitary matrix, 
$U^{\dagger}U=UU^{\dagger}=1$, that replaces the quaternion phase $\Omega$ 
(which is a $1 \times 1$ quaternion unitary matrix) of the preceding section.  
 
Thus, to generalize the results of the preceding section to the non-Abelian 
case (i) one replaces the state vectors $|\psi\rangle, |\hat{\psi}\rangle, 
|\tilde{\psi}\rangle$ by $n$-component column vectors $|\psi_a\rangle, 
|\hat{\psi}_a\rangle, |\tilde{\psi}_a\rangle,~a=1,...,n$, (ii) one replaces 
the phases $\Omega, \hat{\omega}, ...$ by $n \times n$ quaternion unitary 
matrices acting on the state vector indices, (iii) one replaces ${\rm Re}$ in 
Eq.~(22) by ${\rm Re Tr}$, with ${\rm Tr}$ the trace over the subspace 
$V_n$, and (iv) as in Ref.~4, one generalizes the parallel transport condition 
 
of Eq.~(9b) to 
$$\langle \hat{\psi}_a(t) | {\partial | \hat{\psi}_b(t) \rangle 
\over \partial t} 
=0~,~~~a,b=1,...,n.\eqno(28d)$$
The principal difference from the complex 
case treated in Ref.~4 
is that in the quaternion case, the unitary matrix factors must always be 
ordered to the right of ket state vectors, whereas in the complex case the 
ordering is irrelevant, and in fact in Ref.~4 the matrix factors are ordered 
to the left.  The results of Ref.~4 can be obtained by the complex 
specialization of the results obtained in this paper.  However, we have 
introduced here a new technique of using parallel transported states 
$|\hat{\psi}_a \rangle$ to cleanly separate the non-Abelian geometric phase 
and the dynamical phase, which in general (even in the complex non-Abelian 
case) do not commute with each other.  
\vfill\eject
\bigskip
\centerline{\bf 5.~~ Summary and Discussion}
\bigskip
To summarize, we have shown that both the complex Abelian and non-Abelian 
nonadiabatic geometric phases can be generalized to quaternonic Hilbert 
space.  These results are both of theoretical interest, and of experimental 
relevance for possible tests for complex versus quaternionic quantum 
mechanics.  Long ago, Peres [8] proposed testing for quaternionic quantum 
mechanical effects by looking for noncommutativity of scattering phase 
shifts.  However, the result of Ref.~6 that the $S$-matrix in quaternionic 
quantum mechanics is always complex valued (for nonzero energy states) 
implies that there are no 
quaternionic scattering phase shifts, and the Peres test necessarily gives 
a null result.  An alternative but related method is to look for interference 
effects in cyclic evolutions that could show the presence of 
quaternionic effects.  
The fact [6] that the adiabatic geometric phase is always complex (for 
nonzero energy states) is a counterpart of the complexity of the $S$-matrix,  
and implies that a null result will always be obtained for cyclic interference 
experiments involving adiabatic state evolutions.  However, the results 
obtained here show that for cyclic evolutions that are nonadiabatic, one 
could in principle devise interference experiments to place meaningful 
bounds on postulated quaternionic components of the wave function.  
\bigskip
\centerline{\bf Acknowledgments}
\bigskip
The work of SLA was supported in part by the Department of Energy under        
                              
Grant \#DE--FG02--90ER40542, and of JA in part by ONR grant \# R\&T 3124141 
and NSF grant \# PHY-9307708.  SLA wishes to acknowledge the 
hospitality of the Aspen Center for Physics, where part of this work 
was done.  
\vfill\eject
\centerline{\bf References}
\bigskip
\noindent
\item{[1]}  M. V. Berry, Proc. Roy. Soc. London {\bf A 392}, 45 (1984).
\bigskip 
\noindent
\item{[2]}  F. Wilczek and A. Zee, Phys. Rev. Lett. {\bf 52}, 2111 (1984).
\bigskip
\noindent
\item{[3]}  Y. Aharanov and J. Anandan, Phys. Rev. Lett. {\bf 58}, 1593 (1987).
\bigskip
\noindent
\item{[4]}  J. Anandan, Phys. Letters {\bf A 133}, 171 (1988).
\bigskip
\noindent
\item{[5]}  L. P. Horwitz and L. C. Biedenharn, Ann. Phys. {\bf 157}, 432 
(1984).
\bigskip
\noindent
\item{[6]}  S. L. Adler, {\it Quaternionic Quantum Mechanics and Quantum 
Fields}, 
Oxford University Press, New York and Oxford, 1995. 
\bigskip
\noindent
\item{[7]}  P. L\'evay, Phys. Rev. {\bf A 41}, 2837 (1990); J. Math. Phys. 
{\bf 32}, 2347 (1991).
\bigskip
\noindent
\item{[8]} A. Peres, Phys. Rev. Lett. {\bf 42}, 683 (1979).
\vfill
\eject
\bye

Return-Path: adler@sns.ias.edu
Received: from IAS.EDU (mailgate.ias.edu [192.16.204.20]) by 
lonestar.sns.ias.edu (8.6.12/8.6.12) with ESMTP id MAA00430 for 
<val@sns.ias.edu>; Thu, 26 Sep 1996 12:11:16 -0400
Received: from lonestar.sns.ias.edu (lonestar.sns.ias.edu [198.138.243.45]) by 
IAS.EDU (8.6.12/8.6.12) with ESMTP id MAA24217 for <val@ias.edu>; Thu, 26 Sep 
1996 12:09:51 -0400
Received: from nevada.sns.ias.edu (nevada.sns.ias.edu [198.138.243.31]) by 
lonestar.sns.ias.edu (8.6.12/8.6.12) with ESMTP id MAA00426 for <val@ias.edu>; 
Thu, 26 Sep 1996 12:11:15 -0400
From: Stephen Adler <adler@IAS.EDU>
Received: (adler@localhost) by nevada.sns.ias.edu (8.6.12/8.6.12) id MAA24698 
for val@ias.edu; Thu, 26 Sep 1996 12:11:13 -0400
Date: Thu, 26 Sep 1996 12:11:13 -0400
Message-Id: <199609261611.MAA24698@nevada.sns.ias.edu>
To: val@IAS.EDU
Subject: final version for bulletin board and preprint

\input iasmacros
\twelvepoint
\doublespace
{\nopagenumbers{
\rightline{IASSNS-HEP-96/89}
\rightline{~~~September, 1996}
\bigskip\bigskip
\centerline{\rmb Nonadiabatic Geometric Phase}
\centerline{\rmb in Quaternionic Hilbert Space}
\medskip
\centerline{\itb Stephen L. Adler}
\centerline{\bf Institute for Advanced Study}
\centerline{\bf Princeton, NJ 08540}
\centerline{\it adler@sns.ias.edu}
\medskip
\medskip
\centerline{\itb Jeeva Anandan}
\centerline{\bf Department of Physics and Astronomy, University of  
South Carolina}
\centerline{\bf Columbia, South Carolina 29208}
\centerline{\it jeeva@nuc003.psc.sc.edu}
\medskip

\bigskip\bigskip
\leftline{To appear in a special issue of FOUNDATIONS OF PHYSICS dedicated}
\leftline{to Professor Lawrence Horwitz on the occasion of his 65th birthday.}
\bigskip\bigskip
}}
\AB
We develop the theory of the nonadiabatic geometric phase, in both the 
Abelian and non-Abelian cases, in quaternionic Hilbert space.
\AE
\bigskip\bigskip
\vfill\eject
\pageno=2
\centerline{{\bf 1.~~Introduction}}
\bigskip
The theory of geometric phases associated with cyclic evolutions of a 
physical system is now a well-developed subject in complex Hilbert space.
The seminal work of Berry on the adiabatic single state (Abelian) case [1] 
has been extended to the non-Abelian case of the adiabatic evolution of
a set of degenerate states [2], and both of these 
have been further extended [3, 4] 
to show that there is a geometric phase associated with any cyclic but 
nonadiabatic evolution of a single quantum state or of a degenerate group 
of quantum states.  

In this paper we take up another direction for generalization of the 
geometric phase, from quantum mechanics in complex Hilbert 
space to quantum mechanics [5, 6] in quaternionic Hilbert space.  
The generalization of the adiabatic geometric phase to quaternionic Hilbert 
space was given in Ref. 6, where it was shown that for states of nonzero 
energy the adiabatic geometric phase is complex, as opposed to quaternionic, 
with a 
quaternionic adiabatic geometric phase occurring only for the adiabatic 
cyclic evolution of zero energy states.  Consideration of nonadiabatic 
cyclic evolutions was also begun in Ref. 6, but the discussion given there 
is incomplete.  While Sec. 5.8 of Ref.~6 constructed a nonadiabatic cyclic 
invariant phase, it did not address the 
problem of separating this phase into a dynamical part determined by the 
quantum mechanical Hamiltonian, and a geometric part that depends only 
on the ray orbit and is independent of the Hamiltonian.  

The purpose of the present paper is to give a complete discussion of the 
nonadiabatic geometric phase in quaternionic Hilbert space.  In Sec.~2 
we give a very brief survey of the properties of quantum mechanics 
in quaternionic Hilbert space that are needed in the analysis that follows.    
In Sec.~3 we consider the cyclic nonadiabatic evolution of a single quantum 
state, and show how to explictly generalize to quaternionic Hilbert space 
the construction of a nonadiabatic geometric phase given in Ref. 3.  
In Sec.~4 we extend our analysis to the case of a degenerate group of  
states, thereby obtaining a quaternionic nonadiabatic non-Abelian geometric 
phase corresponding to the complex construction given in Ref. 4.
A brief summary and  discussion of our results is given in Sec.~5.  
\bigskip
\centerline{{\bf 2.  Quantum Mechanics in Quaternionic Hilbert Space}}
\bigskip
Only a few properties of quaternionic quantum mechanics are needed for 
the discussion that follows; the reader wishing to learn more than we 
can present here should consult Ref.~6.  In quaternionic quantum mechanics, 
the Dirac transition amplitudes $\langle \psi | \phi \rangle$ are 
quaternion valued, that is, they have the form
$$\langle \psi | \phi \rangle=r_0+r_1 i+r_2 j+r_3 k~,\eqno(1)$$
where $r_{0,1,2,3}$ are real numbers and where $i,j,k$ are quaternion 
imaginary units obeying the associative algebra 
$i^2=j^2=k^2=-1$ and $ij=-ji=k,~jk=-kj=i,~ki=-ik=j$.  Because quaternion 
multiplication is noncommutative, two independent Dirac transition 
amplitudes $\langle \psi | \phi \rangle $ and $\langle \kappa | \eta \rangle$ 
in general do not commute with one another, unlike the situatation in 
standard complex quantum mechanics, where all Dirac transition amplitudes are 
complex numbers and mutually commute.  The transition probability  
corresponding to the amplitude of Eq.~(1) is given by 
$$P(\psi,\phi)=|\langle \psi | \phi \rangle|^2\equiv 
\overline{\langle \psi | \phi \rangle}\langle \psi | \phi \rangle
=r_0^2+r_1^2+r_2^2+r_3^2~,\eqno(2)$$
where the bar denotes the quaternion conjugation operation $\{i,j,k\} \to 
\{-i,-j,-k\}$ and where we have assumed the states $|\psi \rangle$ and 
$|\phi\rangle$ to be unit normalized.  
Since the quaternion norm defined by Eq.~(2) has the multiplicative 
norm property 
$$|q_1 q_2|=|q_1|~|q_2|~,\eqno(3)$$
the transition probability of Eq.~(2) is unchanged when the state vector 
$| \phi \rangle$  is right multiplied by a quaternion $\omega$ of unit  
magnitude, 
$$|\phi\rangle \to |\phi \rangle \omega,~~|\omega|=1~~
\Rightarrow P(\psi,\phi) \to P(\psi, \phi)
~.\eqno(4)$$
Hence as in complex quantum mechanics, physical states are associated with 
Hilbert space rays of the form $\{|\phi\rangle \omega~: |\omega|=1\}$, and 
the transition probability of Eq.~(2) is the same for any ray representative 
state vectors $|\psi \rangle $ and $|\phi \rangle$ chosen from their 
corresponding rays.  In the next section, we shall follow Ref.~3 in denoting 
quaternionic Hilbert space by ${\cal H}$,  and the projective Hilbert 
space of rays of ${\cal H}$ by ${\cal P}$.

Time evolution of the state vector $| \psi \rangle$  is described in 
quaternionic quantum mechanics by the Schr\"odinger equation 
$${\partial |\psi\rangle \over \partial t}=-\tilde H |\psi \rangle~,
\eqno(5a)$$
with 
$$\tilde H=-\tilde H^{\dagger}~\eqno(5b)$$
an anti-self-adjoint Hamiltonian.  From Eqs.~(5a,~b) we see that the 
Dirac transition amplitude $\langle \psi |\phi \rangle$ is time independent, 
$$\eqalign{
{\partial \over \partial t} \langle \psi| \phi \rangle=& 
({\partial \over \partial t} \langle \psi|)| \phi \rangle+ 
\langle \psi| {\partial \over \partial t} | \phi \rangle \cr 
=&\langle \psi| \tilde H - \tilde H |\phi \rangle=0~,\cr
}\eqno(6)$$
and thus the Schr\"odinger dynamics of state vectors 
preserves the inner product structure of Hilbert space.  
The dynamics of Eqs.~(5,~6) is evidently preserved under right linear 
superposition of states with quaternionic constants, 
$$\eqalign{
{\partial |\psi\rangle \over \partial t}=&-\tilde H |\psi \rangle~,
{\partial |\phi\rangle \over \partial t}=-\tilde H |\phi \rangle ~~\Rightarrow 
\cr
{\partial(|\psi\rangle q_1+|\phi\rangle q_2) \over \partial t}=&
-\tilde H (|\psi\rangle q_1+|\phi\rangle q_2)~.\cr
}\eqno(7)$$
Equation (7) illustrates two general features of our conventions for 
quaternionic quantum mechanics,  
which are that linear operators (such as $\tilde H$) act on Hilbert space 
state vectors by 
multiplication from the left, whereas quaternionic numbers (the scalars 
of Hilbert space) act on state vectors by multiplication from the right.   
Adherence to these ordering conventions is essential because of the 
noncommutative nature of quaternionic multiplication.  
\bigskip
\centerline{\bf 3.  The Nonadiabatic Abelian Quaternionic Geometric Phase}
\bigskip
Let us now consider a unit normalized quaternionic Hilbert space 
state $| \psi(t) \rangle$
which undergoes a cyclic evolution between the times $t=0$ and $t=T$.  Since 
physical states are associated with rays, this means that 
$$|\psi(T)\rangle=  |\psi(0) \rangle \Omega~,~~|\Omega|=1~, \eqno(8)$$
and so the orbit ${\cal C}$ of $|\psi(t) \rangle$ in ${\cal H}$ projects to 
a closed curve  $\hat{\cal C}$ in the projective Hilbert space ${\cal P}$.  

Let us now define a state $|\hat {\psi}(t) \rangle$ that is equal to 
$|\psi(t) \rangle$ at $t=0$, that differs from $|\psi(t) \rangle$ only by   
a reraying at general times, i.e.,  
$$\eqalign{
|\psi(t)\rangle=&|\hat{\psi}(t)\rangle \hat{\omega}(t)~,\cr
|\hat{\omega}(t)|=&1~, \cr
\hat{\omega}(0)=1~,\cr
}\eqno(9a)$$
and that evolves in time by parallel transport, 
i.e., 
$$\langle \hat{\psi}(t) | {\partial | \hat{\psi}(t) \rangle 
\over \partial t} 
=0~.\eqno(9b)$$
The conditions of Eqs.~(9a,~b) uniquely determine $\hat{\omega}(t)$, and 
hence the state $|\hat{\psi}(t)\rangle$, as follows.  Substituting 
the first line of Eq.~(9a) into the Schr\"odinger equation of Eq.~(5a), 
we get
$$\eqalign{
-\tilde H |\hat{\psi}(t)\rangle \hat{\omega}(t)=&
-\tilde H |\psi(t)\rangle\cr
=&{\partial |\psi(t)\rangle \over \partial t} 
=|\hat{\psi}(t)\rangle {d \hat{\omega}(t) \over d t}
+{\partial |\hat{\psi}(t)\rangle \over \partial t} \hat{\omega}(t)~.\cr
}\eqno(10)$$
Taking the inner product of this equation with the state $\langle \hat{\psi}
(t)|$, and using the 
unit normalization of the state vector $|\hat{\psi}(t)\rangle$ together
with the parallel transport condition of Eq.~(9b), we get
$${d \hat{\omega}(t) \over d t}=
-\langle \hat{\psi}(t)|\tilde H|\hat{\psi}(t)\rangle 
\hat{\omega}(t)~.\eqno(11)$$
This differential equation can be immediately integrated to give 
$$\hat{\omega}(t)=T_{\ell}e^{-\int_0^t dv \langle \hat{\psi}(v)|\tilde H
|\hat{\psi}(v) \rangle}~,\eqno(12)$$ 
where $T_{\ell}$ denotes the time ordered product which orders later times to 
the 
left, and where we have used the initial condition on the third line of 
Eq.~(9a).  In particular, Eq.~(12) gives us a formula for the value 
$\hat{\omega}(T)$ at the end of the cyclic evolution.  We shall see that 
this has the interpretation of the dynamics-dependent part of the total 
phase change $\Omega$.

To relate Eq.~(12) to the total phase change, we use Eqs.~(8) and (9a) to 
write 
$$|\hat{\psi}(T)\rangle \hat{\omega}(T)= |\psi(T) \rangle=
|\psi(0) \rangle \Omega=|\hat{\psi}(0)\rangle \Omega~,\eqno(13a)$$
so that taking the inner product with $\langle \hat{\psi}(0)|$ gives
$$\Omega=\langle \hat{\psi}(0)|\hat{\psi}(T) \rangle 
\hat{\omega}(T)~.\eqno(13b)$$
To complete the calculation, we must now evaluate the inner product 
appearing in Eq.~(13b).  To do this, we introduce a third state vector 
$|\tilde{\psi}(t)\rangle$ which differs from $|\hat{\psi}(t)\rangle$ by a 
change of ray representative, by writing
$$\eqalign{
|\hat{\psi}(t)\rangle =& |\tilde{\psi}(t)\rangle \tilde{\omega}(t)~,\cr
|\tilde{\omega}(t)|=&1~,\cr
\tilde{\omega}(0)=&1~,\cr
}\eqno(14a)$$
and by requiring that $\tilde{\psi}$ should be continuous over the orbit 
${\cal C}$, 
$$|\tilde{\psi}(T)\rangle=|\tilde{\psi}(0)\rangle~.\eqno(14b)$$
Differentiating the first line of Eq.~(14a) with respect to time, we get 
$${\partial |\hat{\psi}(t)\rangle \over \partial t}=
{\partial |\tilde{\psi}(t)\rangle \over \partial t} \tilde{\omega}(t)
+|\tilde{\psi}(t)\rangle {d \tilde{\omega}(t) \over d t}~.
\eqno(15)$$
Taking the inner product of Eq.~(15) with 
$\tilde {\omega}(t)\langle \hat{\psi}(t)|$, 
using the parallel transport condition of Eq.~(9b) together with the 
first line of Eq.~(14a), and abbreviating the time derivative ${\partial \over 
\partial t}$ by a dot, we obtain
$$0=\langle \tilde{\psi}(t)| \dot{\tilde{\psi}}(t) \rangle \tilde{\omega}(t)
+\langle \tilde{\psi}(t)|\tilde{\psi}(t)\rangle \dot{\tilde{\omega}}(t)~.
\eqno(16a)$$
Since the second line of Eq.~(14a) implies that the state $|\tilde{\psi}(t)
\rangle$ is unit normalized, Eq.~(16a) simplifies to 
$$\dot{\tilde{\omega}}(t)=
-\langle \tilde{\psi}(t)| \dot{\tilde{\psi}}(t) \rangle \tilde{\omega}(t)~,
\eqno(16b)$$
which can be immediately integrated to give 
$$\tilde{\omega}(t)=T_{\ell}e^{-\int_0^t dv \langle \tilde{\psi}(v)|
\dot{\tilde{\psi}}(v) \rangle}~,\eqno(17)$$ 
with $T_{\ell}$ as before indicating a time ordered product.  In particular, 
Eq.~(17) gives us a formula for $\tilde{\omega}(T)$.  But from Eqs.~(14a, b) 
we have
$$|\hat{\psi}(T)\rangle=|\tilde{\psi}(T)\rangle \tilde{\omega}(T)=
|\tilde{\psi}(0)\rangle \tilde{\omega}(T)=|\hat{\psi}(0)\rangle 
\tilde{\omega}(T)~,\eqno(18a)$$
and so taking the inner product of Eq.~(18a) with $\langle \hat{\psi}(0)|$
we get 
$$\langle \hat{\psi}(0)|\hat{\psi}(T)\rangle = \tilde{\omega}(T)~,\eqno(18b)$$
determining the inner product appearing in Eq.~(13b).  

We thus get as our final result, 
$$\Omega=\Omega_{\rm geometric} \Omega_{\rm dynamical}~,\eqno(19a)$$
with 
$$\Omega_{\rm geometric}\equiv\tilde{\omega}(T)=
T_{\ell}e^{-\int_0^T dv \langle \tilde{\psi}(v)|
\dot{\tilde{\psi}}(v) \rangle}~,\eqno(19b)$$ 
and with
$$\Omega_{\rm dynamical}\equiv\hat{\omega}(T)=
T_{\ell}e^{-\int_0^T dv \langle \hat{\psi}(v)|\tilde H
|\hat{\psi}(v) \rangle}~.\eqno(19c)$$ 
The dynamical part of the phase is so called because it depends explicitly  
on $\tilde H$, as well as on the orbit $\hat{\cal C}$ in the projective 
Hilbert space $\cal P$; it is uniquely determined   
by the conditions of Eqs.~(9a, b), since these conditions uniquely   
determine the state $|\hat{\psi}(t)\rangle$.
The geometric part of the phase is so called because, as we shall now 
show, it depends uniquely on the projective orbit $\hat{\cal C}$ 
up to an overall quaternion automorphism transformation.  To see this, 
let us make the reraying 
$$|\tilde{\psi}(t)\rangle \to  |\tilde{\psi}^{\,\prime}\rangle 
\omega^{\,\prime}
(t),~~|\omega^{\,\prime}|=1~, \eqno(20a)$$
with $\omega^{\,\prime}(t)$ continuous over the orbit $\cal C$ so that 
$$\omega^{\,\prime}(T)=\omega^{\,\prime}(0)~.\eqno(20b)$$
Then (as shown in detail in Sec. 5.8 of Ref.~6) 
the properties of the time ordered integral in Eq.~(19b) imply that under 
this transformation, 
$$\Omega_{\rm geometric} \to \overline{\omega}^{\,\prime}(T) \Omega_{\rm 
geometric}
\,\omega^{\,\prime}(0)~,\eqno(21a)$$
which by the continuity condition of Eq.~(20b) reduces to the 
quaternion automorphism transformation 
$$\Omega_{\rm geometric} \to \overline{\omega}^{\,\prime}(0) \Omega_{\rm 
geometric}
\,\omega^{\,\prime}(0)~.\eqno(21b)$$
Since for any two quaternions $q_1, q_2$ we have ${\rm Re}\, q_1 q_2=
{\rm Re}\, q_2 q_1$, with Re denoting the real part, Eq.~(21b) implies 
that 
$$\cos \gamma_{\rm geometric} \equiv {\rm Re} \Omega_{\rm 
geometric}~\eqno(22)$$ 
is a reraying invariant, and thus $\gamma_{\rm geometric}$ is a nonadiabatic 
geometric  phase angle that is a property solely of the projective 
orbit $\hat{\cal C}$.  The fact that the nonadiabatic geometric phase 
in quaternionic Hilbert space is only determined modulo $\pi$ is a reflection 
of the fact that $e^{i\gamma}$ is changed to $e^{-i\gamma}$ by 
the quaternion automorphism transformation 
$$e^{-i\gamma}=\bar j e^{i \gamma} j ~.\eqno(23)$$
Thus, to recover the result that the complex nonadiabatic geometric phase is 
determined modulo $2 \pi$ by embedding a complex Hilbert space in a 
quaternionic one and using Eqs.~(19a-c), one must exclude the possibility 
of making intrinsically quaternionic automorphism transformations involving 
the quaternion units $j$ or $k$, as in Eq.~(23).

In geometric terms, $\Omega_{\rm geometric}$ is the holonomy transformation 
of the connection $A \equiv \langle \tilde \psi | d \tilde \psi \rangle$.  
But since this connection is quaternion-imaginary valued, it is analogous 
to an $SO(3)$ gauge potential.  Therefore, the corresponding curvature 
is of the Yang-Mills type and is given by $F=dA + A \wedge A$.
  
An alternative expression for the total phase change $\Omega$ can be 
obtained [7] by writing
$$\eqalign{
|\psi(t)\rangle=&|\tilde{\psi}(t)\rangle \tilde {\chi}(t)~,\cr
\tilde{\chi}(t)=&\hat{\omega}(t)\tilde{\omega}(t)~,~~\tilde{\chi}(0)=1~.\cr
}\eqno(24)$$
Substituting Eq.~(24) into the Schr\"odinger equation and then 
taking the inner product with $\langle \tilde{\psi}(t)|$, we obtain
$${d \tilde{\chi}(t) \over dt}=-(\langle \tilde{\psi}(t) | \tilde H | 
\tilde{\psi}(t) \rangle + \langle \tilde{\psi}(t)|\dot{\tilde{\psi}}(t)
\rangle )\tilde{\chi}(t)~, \eqno(25a)$$
which can be integrated from $0$ to $T$ to give 
$$\Omega=T_{\ell}e^{-\int_0^T dv
(\langle \tilde{\psi}(v) | \tilde H | 
\tilde{\psi}(v) \rangle + \langle \tilde{\psi}(v)|\dot{\tilde{\psi}}(v)
\rangle)}~.\eqno(25b)$$
This procedure and the resulting formula of Eq.~(25b) are direct analogs 
of the derivation given in Ref.~3 for the complex Hilbert space case, but in 
quaternionic 
Hilbert space the two terms in the exponential are noncommutative, and so 
the exponential in Eq.~(25b) cannot be immediately factored into
dynamical and geometric phase factors.  As we have seen, to achieve this 
factorization it is necessary to use a two-step procedure, involving 
the parallel transported state $|\hat{\psi}(t)\rangle$ as well as the 
state $|\tilde{\psi}(t)\rangle$ that is continuous over the cycle.
\bigskip
\centerline{\bf 4.  The Nonadiabatic Non-Abelian Quaternionic Geometric Phase}
\bigskip                                                                 
We turn next to the quaternionic Hilbert space generalization of the 
complex nonadiabatic [4] non-Abelian [2] geometric phase.  We consider 
now a cyclic evolution in a $n$-dimensional Hilbert subspace $V_n$, i.e., 
$V_n(T)=V_n(0)$.  Let $|\psi_a(t)\rangle,~a=1,...,n$ be a complete orthonormal 
basis for $V_n$, so that the reraying invariant 
projection operator for $V_n$ is 
$$\rho_n(t)=\sum_{a=1}^n |\psi_a(t)\rangle \langle \psi_a(t)|~,\eqno(26a)$$
in terms of which the cyclic evolution condition takes the 
form 
$$\rho_n(T)=\rho_n(0)~.\eqno(26b)$$
Expressed in terms of the state vectors of the basis, the invariance of $V_n$
implies that the basis element $|\psi_b(T)\rangle$ must be a superposition  
of the basis elements $|\psi_a(0)\rangle$, multiplied from the right 
by quaternionic coefficients $U_{ab}$, 
$$|\psi_b(T)\rangle = \sum_{a=1}^n |\psi_a(0)\rangle U_{ab}~. \eqno(27)$$
Substituting Eq.~(27) into Eqs.~(26a, b), we find that invariance of the 
projection operator requires  
$$\eqalign{
\rho_n(T)=&\sum_{b=1}^n|\psi_b(T)\rangle \langle \psi_b(T)|  \cr
=&\sum_{a,b,c=1}^n |\psi_a(0)\rangle U_{ab} \overline{U}_{cb} \langle 
\psi_c(0)| \cr
=&\sum_{a=1}^n |\psi_a(0)\rangle \langle \psi_a(0)| = \rho_n(0)~, \cr
}\eqno(28a)$$
which implies that 
$$\sum_{b=1}^n U_{ab} \overline{U}_{cb}=\delta_{ac}~.\eqno(28b)$$
Similarly, orthonormality of the basis at $t=0$ and $t=T$ implies that 
$$\eqalign{
\delta_{ab}=&\langle \psi_a(T)|\psi_b(T) \rangle =\sum_{c,d=1}^n 
\overline{U}_{ca}\langle \psi_c(0)|\psi_d(0) \rangle U_{db} \cr  
=&\sum_{c,d=1}^n \overline{U}_{ca} \delta_{cd} U_{db}   
=\sum_{d=1}^n    \overline{U}_{da}  U_{db} ~.\cr
}\eqno(28c)$$
Hence $U$ is a $n \times n$ quaternion unitary matrix, 
$U^{\dagger}U=UU^{\dagger}=1$, that replaces the quaternion phase $\Omega$ 
(which is a $1 \times 1$ quaternion unitary matrix) of the preceding section.  
 
Thus, to generalize the results of the preceding section to the non-Abelian 
case (i) one replaces the state vectors $|\psi\rangle, |\hat{\psi}\rangle, 
|\tilde{\psi}\rangle$ by $n$-component column vectors $|\psi_a\rangle, 
|\hat{\psi}_a\rangle, |\tilde{\psi}_a\rangle,~a=1,...,n$, (ii) one replaces 
the phases $\Omega, \hat{\omega}, ...$ by $n \times n$ quaternion unitary 
matrices acting on the state vector indices, (iii) one replaces ${\rm Re}$ in 
Eq.~(22) by ${\rm Re Tr}$, with ${\rm Tr}$ the trace over the subspace 
$V_n$, and (iv) as in Ref.~4, one generalizes the parallel transport condition 
 
of Eq.~(9b) to 
$$\langle \hat{\psi}_a(t) | {\partial | \hat{\psi}_b(t) \rangle 
\over \partial t} 
=0~,~~~a,b=1,...,n.\eqno(28d)$$
The principal difference from the complex 
case treated in Ref.~4 
is that in the quaternion case, the unitary matrix factors must always be 
ordered to the right of ket state vectors, whereas in the complex case the 
ordering is irrelevant, and in fact in Ref.~4 the matrix factors are ordered 
to the left.  The results of Ref.~4 can be obtained by the complex 
specialization of the results obtained in this paper.  However, we have 
introduced here a new technique of using parallel transported states 
$|\hat{\psi}_a \rangle$ to cleanly separate the non-Abelian geometric phase 
and the dynamical phase, which in general (even in the complex non-Abelian 
case) do not commute with each other.  
\vfill\eject
\bigskip
\centerline{\bf 5.~~ Summary and Discussion}
\bigskip
To summarize, we have shown that both the complex Abelian and non-Abelian 
nonadiabatic geometric phases can be generalized to quaternonic Hilbert 
space.  These results are both of theoretical interest, and of experimental 
relevance for possible tests for complex versus quaternionic quantum 
mechanics.  Long ago, Peres [8] proposed testing for quaternionic quantum 
mechanical effects by looking for noncommutativity of scattering phase 
shifts.  However, the result of Ref.~6 that the $S$-matrix in quaternionic 
quantum mechanics is always complex valued (for nonzero energy states) 
implies that there are no 
quaternionic scattering phase shifts, and the Peres test necessarily gives 
a null result.  An alternative but related method is to look for interference 
effects in cyclic evolutions that could show the presence of 
quaternionic effects.  
The fact [6] that the adiabatic geometric phase is always complex (for 
nonzero energy states) is a counterpart of the complexity of the $S$-matrix,  
and implies that a null result will always be obtained for cyclic interference 
experiments involving adiabatic state evolutions.  However, the results 
obtained here show that for cyclic evolutions that are nonadiabatic, one 
could in principle devise interference experiments to place meaningful 
bounds on postulated quaternionic components of the wave function.  
\bigskip
\centerline{\bf Acknowledgments}
\bigskip
The work of SLA was supported in part by the Department of Energy under        
                              
Grant \#DE--FG02--90ER40542, and of JA in part by ONR grant \# R\&T 3124141 
and NSF grant \# PHY-9307708.  SLA wishes to acknowledge the 
hospitality of the Aspen Center for Physics, where part of this work 
was done.  
\vfill\eject
\centerline{\bf References}
\bigskip
\noindent
\item{[1]}  M. V. Berry, Proc. Roy. Soc. London {\bf A 392}, 45 (1984).
\bigskip 
\noindent
\item{[2]}  F. Wilczek and A. Zee, Phys. Rev. Lett. {\bf 52}, 2111 (1984).
\bigskip
\noindent
\item{[3]}  Y. Aharanov and J. Anandan, Phys. Rev. Lett. {\bf 58}, 1593 (1987).
\bigskip
\noindent
\item{[4]}  J. Anandan, Phys. Letters {\bf A 133}, 171 (1988).
\bigskip
\noindent
\item{[5]}  L. P. Horwitz and L. C. Biedenharn, Ann. Phys. {\bf 157}, 432 
(1984).
\bigskip
\noindent
\item{[6]}  S. L. Adler, {\it Quaternionic Quantum Mechanics and Quantum 
Fields}, 
Oxford University Press, New York and Oxford, 1995. 
\bigskip
\noindent
\item{[7]}  P. L\'evay, Phys. Rev. {\bf A 41}, 2837 (1990); J. Math. Phys. 
{\bf 32}, 2347 (1991).
\bigskip
\noindent
\item{[8]} A. Peres, Phys. Rev. Lett. {\bf 42}, 683 (1979).
\vfill
\eject
\bye